\documentclass{article}
\usepackage{amsmath}

\usepackage{graphicx}

\begin{document}

\title{Delayed Dynamics with Transient Resonating Oscillations}


\author{Kenta Ohira$^{1}$ and Toru Ohira$^{2}$\\
\ $^{1}$Future Value Creation Research Center,\\ Graduate School of Informatics, Nagoya University, Japan\\
\ $^{2}$Graduate School of Mathematics, Nagoya University, Japan
}

\maketitle

\begin{abstract}
Recently, we have studied a delay differential equation which has a coefficient that is a linear function of time. 
The equation has shown the oscillatory transient dynamics appear and disappear as the delay is increased between zero to asymptotically large delay. We here propose and study another equation that shows similar transient oscillations. It has an extra exponential gaussian factor on the delayed feedback term. It is shown that this equation is analytically tractable with the use of the Lambert $W$ function. This equation is also studied numerically to confirm some of the properties inferred from the analytical solution. We also have found that the amplitude of transient oscillation changes and goes through a maximum as we increase the value of the delay. In this sense, the proposed equation is one of the simplest dynamical equations that brings out a resonant behavior without any external oscillating inputs. 
\end{abstract}


\section{Introduction}
Delays exist in many
control and mutually interacting systems and have been investigated
in various fields including mathematics, biology, physics, engineering, and economics.\cite{heiden1979,bellman1963,cabrera1,hayes1950,insperger,kcuhler,longtinmilton1989a,mackeyglass1977,miltonetal2009b,ohirayamane2000,smith2010,stepan1989,stepaninsperger,szydlowski2010}). 
Typically,  delays cause instability of stable fixed points leading to
oscillatory and more complex dynamics. A representative example is the Mackey--Glass equation\cite{mackeyglass1977}, which shows the sequence of the monotonic convergence, transient oscillations, persistent oscillations, and chaotic dynamics with increasing feedback delay. The path to the complex behaviors of many systems with delays, including this model, is a difficult subject, and 
understanding has been gradually gained (e.g.\cite{taylor}). There is, however,  more to be explored, particularly concerning the nature of time-dependent dynamical trajectories.

``Delay Differential Equations (DDE)'' are the main mathematical approaches and modeling tools for such systems. Typically, DDEs with constant coefficients have been investigated. Recently, we have studied a DDE which has a coefficient that is a linear function of time\cite{kentaohira2021}. Even though this is formally a small change in the equation, 
it shows the oscillatory transient dynamics appear and disappear as the delay is increased between zero to an asymptotically large delay. Also, a new type of resonating behavior has been observed contrasting with the constant-coefficient case.

Here, we propose and study another equation that shows similar transient oscillations in this paper. It has an extra exponential gaussian factor on the delayed feedback term. Though it makes this equation more complex, we will show that it is analytically more tractable with the use of the Lambert $W$ function. In particular, we derive a semi-analytical approximation to capture the dynamical behaviors with varying delays. This equation is then studied numerically to compare with the approximation. The approximation works reasonably well for a range of delay values. Also, we have found that the amplitude of transient oscillation changes and goes through a maximum as we increase the value of the delay. The value of optimal delay also can be estimated by the usage of the $W$ function. In this sense, the proposed equation is one of the simplest dynamical equations that brings out a resonant behavior without external oscillating inputs. In addition, we confirm some of the properties inferred from the analytical investigations. 

It should be noted that we are not investigating stability-switching phenomena (e.g.\cite{yan}) with the delay as the bifurcation parameter. Indeed, in our analysis of the proposed model in this and previous work\cite{kentaohira2021}, the asymptotic stability of the fixed point never changes with increasing delay. 
Instead, the shapes of dynamical trajectories approaching the stable fixed point change with the above-mentioned resonant phenomena.

We close the paper with a brief discussion of
transient oscillations from delay differential equations of similar types.

\section{Main equation and its properties}
The general form of the equation we are interested in is given by the following.
\begin{equation}
{dX(t)\over dt} + a t X(t) = f(t, X(t), X(t-\tau))
\label{dr}
\end{equation}
where $a \geq 0$, $\tau \geq 0$ are real parameters,  and $\tau$ is interpreted as a delay.
This is a slight extension of the constant-coefficient linear delay differential equation describing the dynamics of the variable $X(t)$. The notable difference is that we have $a t$ instead of $a$ in the second term of the equation. Though this appears to be a small change, the behavior of $X(t)$ becomes quite different. In the special case, the function $f$ is identically zero, this can be viewed as the equation for the ground state of the quantum simple harmonic oscillator with the interpretation of $t$ as a position rather than time (e.g.\cite{sakurai}).

For example, we have proposed and studied the following simple special case,
\begin{equation}
{dX(t)\over dt} + a t X(t) = b X(t-\tau)
\label{dr}
\end{equation}
so that the right-hand side is a simple linear function of $X(t-\tau)$ with a constant real parameter coefficient $b$. This is a simple modification of the much-studied Hayes's equation, a first-order delay differential equation with constant coefficients\cite{hayes1950}. We have shown, however, that its behavior is quite different. It gives rise to the behavior where oscillatory transient dynamics appear and disappear as the value of delay increases. The resonant phenomena with respect to the delay have also been observed.

In this paper, we extend the above equation (\ref{dr}) as
\begin{equation}
{dX(t)\over dt} + a t X(t) = b e^{- a \tau t} X(t-\tau).
\label{dr2}
\end{equation}
With the exponent factor inserted in the right-hand side of the equation, it appears more
complex to analyze. We, nevertheless, investigate this equation and show that previous analytical knowledge using the Lambert 
$W$ functions can be employed. 

\subsection{Analysis}

We start with the case that $b=0$. With the initial condition $X(t=0) = X_0$, the solution to the equation is given as
\begin{equation}
X(t) = X_0 e^{- {1\over 2}a t^2}
\label{sho}
\end{equation}
Thus, its dynamics have a trajectory with a gaussian shape. Also, if we consider $t$ as a position rather than time, this is the ground state of the quantum simple harmonic oscillator.

On the other hand, the case that $a=0$ becomes 
\begin{equation}
{dX(t)\over dt}  = b X(t-\tau).
\label{hyen}
\end{equation}
This is the simplest first-order delay differential equation with constant coefficients and  is a special case of the Hayes equation 
\cite{hayes1950}. 

This equation has been much studied and we know the following.

\begin{itemize}
\item
It is known that the dynamics of $X(t)$ monotonically approach to the asymptotically stable origin $X=0$ in the range of
\begin{equation}
0 > b > - {1/{e \tau} }.
\label{c1}
\end{equation}
With $b < - {1/{e \tau} }$, the oscillatory dynamics begin to appear.

\item
Including the above, in the range of 
\begin{equation}
0 > b > - {\pi/{2 \tau} },
\label{c2}
\end{equation}
$X=0$ is asymptotically stable.

\item
Hence, for $b<0 $, the critical delay $\tau_c$ for the loss of the stability of the origin $X=0$ is 
\begin{equation}
\tau_c = - {\pi/{2 b} }.
\label{cdelay}
\end{equation}
At this point, $X(t)$ has a stationary sinusoidal solution with constant amplitude with the angular frequency $\omega_c = |b|$. Equivalently, the critical period of the oscillation is 
\begin{equation}
T_c = 2\pi/\omega_c = 4\tau_c. 
\label{cperiod}
\end{equation}

\item
Further, the general solution of (\ref{hyen}) can be expressed by using the Lambert $W$ function\cite{shinozaki,pusenjak2017}, which is defined
as a multivalued complex function $W: C \rightarrow C$ satisfying
\begin{equation}
W(z)e^{W(z)} = z
\label{wf}
\end{equation}
The branches of the $W$ function are expressed as $W_k, k=0, \pm 1, \pm 2, \dots, \pm \infty$.
Using this function, the general solutions can be written as the following.
\begin{equation}
X(t) = \sum_{k = -\infty}^{\infty} C_k e^{\lambda_k t}, \quad \lambda_k = {1 \over \tau} W_k (b \tau).
\label{solwf1}
\end{equation}
We note that $\lambda_k$ are the roots of the transcendental characteristic equation of (\ref{hyen}),
\begin{equation}
\lambda  = b e^{-\tau\lambda},
\label{chara}
\end{equation}
and that $C_k$ are the constant coefficients determined by the initial interval condition $X(t) = \phi(t) , [-\tau, 0]$ of (\ref{hyen}).
\end{itemize}

The general solution of  (\ref{dr2}) can now be obtained by combining the above two cases. Namely, we set
\begin{equation}
X(t) =  e^{- {1\over 2}a t^2}\hat{X}(t) ,
\label{tau0g}
\end{equation}
then, it is easy to show that $\hat{X}(t)$ satisfies the same form of the differential equation as (\ref{hyen}):
\begin{equation}
{d\hat{X}(t)\over dt}  = \hat{b} \hat{X}(t-\tau), \quad \hat{b} =  \hat{b}(\tau) = b e^{- {1\over 2}a {\tau}^2}
\label{hyen2}
\end{equation}
This leads to the general solution of (\ref{dr2}) using (\ref{solwf1}) and (\ref{tau0g}) as
\begin{equation}
X(t) = e^{- {1\over 2}a t^2} \sum_{k = -\infty}^{\infty} C_k e^{{1 \over \tau} W_k (\hat{b} \tau) t}.
\label{solwf3}
\end{equation}
\clearpage

We can infer qualitatively some properties of this solution.

\begin{itemize}
\item
The first gaussian factor dominates as $t \rightarrow \infty$. Thus, for $a>0$, the asymptotic stability of the origin is kept regardless of the value of the delay.

\item
The oscillatory behavior of the solution arises due to the second factor. Thus, as we have mentioned, the dynamics of $X(t)$ monotonically approach to the asymptotically stable origin $X=0$ in the range of
\begin{equation}
0 > \hat{b}(\tau)\tau > - {1/{e} }.
\label{c1b}
\end{equation}
With $\hat{b}(\tau)\tau < - {1/{e}}$, the oscillatory dynamics begin to appear.
\end{itemize}

In the next subsection, we first investigate the approximate dynamics from equation (\ref{dr2}) using a formal solution (\ref{solwf3}) .

\subsection{Approximate solutions}

We utilize the formal form (\ref{solwf3}) of the solution of equation (\ref{dr2}) in order to approximate its dynamics. The tabulated values of the $W$ function (Appendix) are employed as well as some properties of the $W$ function. We argue that 
the formal solution (\ref{solwf3}) can be approximated  for the ranges of parameters we are showing here by the following.

\begin{equation}
X(t) \approx X_0 e^{- {1\over 2}a t^2} Re[e^{{1 \over \tau} W_0 (\hat{b} \tau) t}] = X_0 e^{- {1\over 2}a t^2} e^{{1 \over \tau} Re[W_0 (\hat{b} \tau)] t}
\cos({1 \over \tau} Im[W_0 (\hat{b} \tau)] t)
\label{solwf4}
\end{equation}

This means we are approximating the solution by a single term using the $0$th (principal) branch of the $W$ function instead of the sum over different branch terms. This is based on the following properties of the $W$ function proved by Shinozaki and Mori\cite{shinozaki}.

When $x$ is a real value, the maximum value of the real part of the $W$ function, $Re[W_k(x)]$, is obtained when 

(i) $k=0$, for $x > -1/e$ 

(ii) $k=0$ and $k = -1$, for $x \leq -1/e$ with $Re[W_0(x)] = Re[W_{-1}(x)]$
In other words, the maximum values are the following for the real value $x$:

(i)  $Re[W_0(x)]$, ($x > -1/e$)

(ii)  $Re[W_0(x)] = Re[W_{-1}(x)]$, ($x \leq -1/e$)

\noindent
Thus, the dominant term in the summation in (\ref{solwf3}) is either (i) $k=0$ or (ii) $k = 0$ and $-1$ (see Appendix).

Another assumption we make is the following. The coefficients $C_k$ in the sum of (\ref{solwf3}) need to
be determined by the initial condition function $\{\phi(t)| -\tau < t \leq 0\}$. We, however, require the matching only at $t=0$ for our case of constant initial condition function $X_0$. 

Putting these rather physical assumptions and the above properties together we obtain  the approximation
(\ref{solwf4}) for the solution of (\ref{dr2}). Note that the case of (ii) with two terms $k = 0$ and $-1$ can also reduce to (\ref{solwf4}) by the properties that $Re[W_0(x)] = Re[W_{-1}(x)]$ and that $Im[W_0(x)] = - Im[W_{-1}(x)]$ which we have confirmed numerically in the Appendix.

The approximate solution also suggests transient oscillations due to the cosine factor.
We can also use the $W$ function to estimate the resonant point. From the structure of our approximation
(\ref{solwf4}), we can infer that the maximum amplitude of the oscillation as we vary the delay can be obtained when the value of the exponent $g(\tau)\equiv {1 \over \tau} Re[W_0 (\hat{b}(\tau) \tau)]$ is positive and the largest. Thus, studying the function $g(\tau)$ can lead us to the optimal delay for the resonant point.

\section{Comparison with numerical simulations}

We are now in the position of studying equation (\ref{dr2}) numerically for comparison against our approximations and confirming some of the characteristics. As mentioned, the oscillatory behaviors appear and disappear as we increase the value of the delay $\tau$, which can be considered a resonating phenomenon. 

The typical dynamics for the case of $a>0, b<0$ are shown in Figure 1. As we noted in the previous section, the asymptotic stability of the origin is kept even for large delays. We compare these with the approximate solution given by (\ref{solwf4}) using the tabulated numerical values for the $W$ function. As shown in Figure 2, the agreement with numerical results is fairly reasonable by this semi-analytical approximation. This approximation works less well with larger delays (e.g. Figure 2(H)) as more branch values $Re[W_k (\hat{b}(\tau) \tau)]$ of the $W$ function become positive with increasing delays. (see Appendix).  

Also, we note that the amplitude of the oscillation changes and  goes through the maximum as we increase the value of the delay (Figure 3(A)). In this sense, we have resonant phenomena with the delay as a tuning parameter. We can again use our approximate solution (\ref{solwf4}) to estimate the resonant point at which the oscillation amplitude is maximal as we change the delay. As mentioned, this can be achieved by plotting the factor in the solution
\begin{equation}
g(\tau)\equiv {1 \over \tau} Re[W_0 (\hat{b}(\tau) \tau)]
\label{resoW}
\end{equation}
and finding the positive maximum point numerically. It is given in Figure 3(B). The estimated resonant point is at $\tau \approx  2.13$, which agrees reasonably well with the one seen in Figure 3(A).

Thus, equation (\ref{dr2}) is one of the simplest dynamical equations showing a resonance without any external oscillatory inputs. These properties are in contrast to the case of (\ref{hyen}), where the stability of the origin is lost beyond the critical delay, and the amplitude of the oscillation is a monotonic increasing function of the delay.

On the other hand, some of the other properties are shared with the case of (\ref{hyen}).
The results of numerical simulations are shown in Figure 4 and Figure 5. In Figure 4, the beginning of the non-monotonic approach to the stable origin appears approximately at the delay $\tau_o$ that satisfies $\hat{b}(\tau_o)\tau_o = - {1/{e}}$. 
This can be also inferred from our approximate solution with the $W$ function (\ref{solwf4}). This is because the imaginary part $Im[W_0 (b \tau)]$ has
the following property
\begin{equation}
Im[W_0 (\hat{b}(\tau)\tau)]\begin{cases}
                    = 0 & (\hat{b}(\tau)\tau \leq  - {1/{e}})\\
    1/b               >  0 & (\hat{b}(\tau) \tau  >  - {1/{e}})
\end{cases}
\label{imWprop}
\end{equation}

In Figure 5, we showed the period of oscillation as a function of $\tau$. The period of oscillation $T_c$ near the critical delay, $\tau_c$, for the case of (\ref{hyen}) is approximately $4 \tau_c$ consistent with (\ref{cperiod}). In the case of (\ref{dr2}), the same argument gives the critical delay $\tau_c$ satisfies 
$\hat{b}(\tau_c)\tau_c = - {\pi/{2}}$. The numerical simulations shows the same property, $T_c \approx 4 \tau_c$.
Again, this can be also inferred from the cosine term of (\ref{solwf4}) together with the property that $Im[W_0 (- {\pi/{2}})] = {\pi/{2}}$. More generally, we can obtain that the period $T$ of oscillation of 
(\ref{dr2}) with varying $\tau$ is approximated by the following 
\begin{equation}
T = { 2\pi\tau \over  Im[W_0 ( \hat{b}(\tau)\tau)]}.
\label{Wperiod}
\end{equation}

\begin{figure}[h]
\begin{center}
\includegraphics[height=12.4cm]{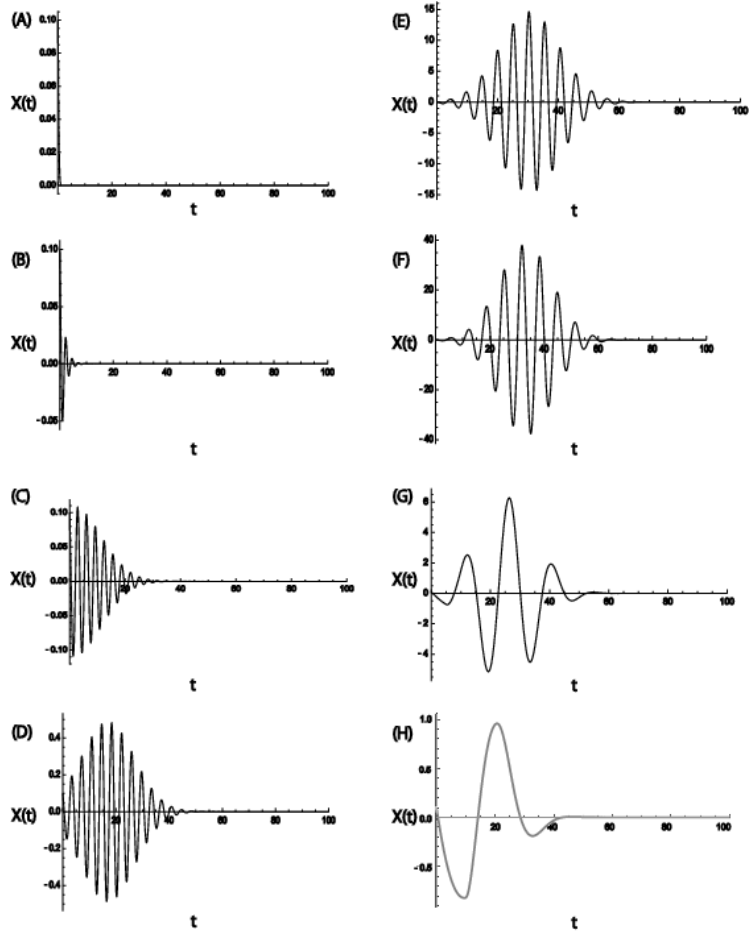}
\caption{Representative dynamics of the main equation (3) with different values of the delays, $\tau$. The parameters are set at $a=0.01, b= - 2.0$ with the initial interval condition as
$X(t) = 0.1 (-\tau \leq t \leq 0)$. The values of the delays $\tau$ are (A)$0.2$, (B)$0.5$, (C)$0.8$, (D)$1.0$, (E)$1.5$, (F)$2.0$, (G)$5$, (H)$10$.}
\label{dynamics}
\end{center}
\end{figure}
\clearpage

\begin{figure}[h]
\begin{center}
\includegraphics[height=12.4cm]{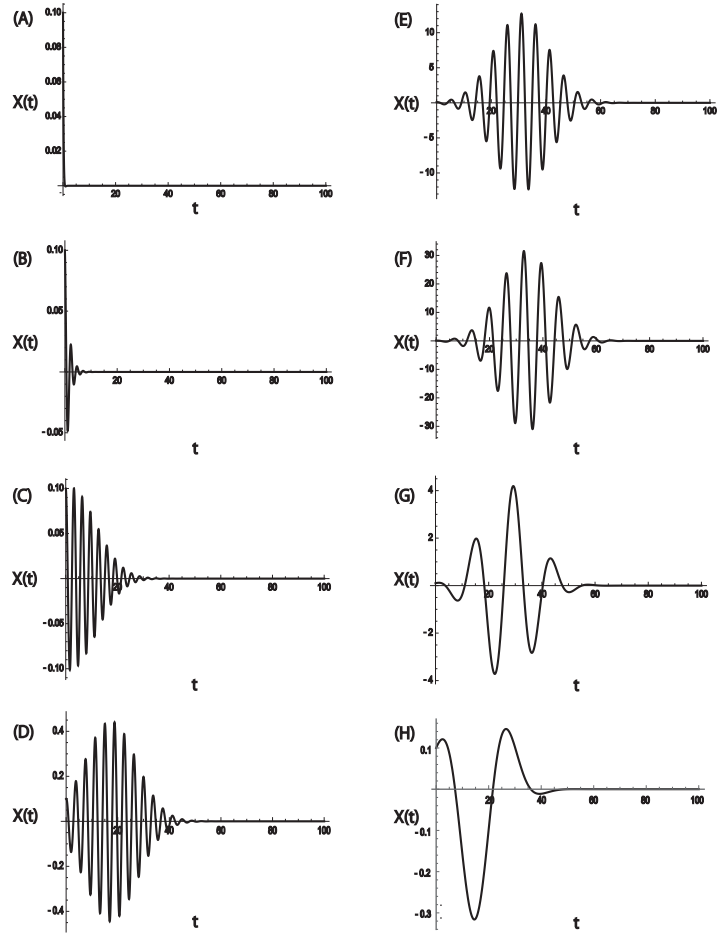}
\caption{Representative plots of the approximate solution  (\ref{solwf4}) with different values of the delays, $\tau$. The parameters are set at the same as in Figure 1: $a=0.01, b= - 2.0$ with the initial interval condition as
$X(t) = 0.1 (-\tau \leq t \leq 0)$. The values of the delays $\tau$ are also the same: (A)$0.2$, (B)$0.5$, (C)$0.8$, (D)$1.0$, (E)$1.5$, (F)$2.0$, (G)$5$, (H)$10$.}
\label{dynamicsW}
\end{center}
\end{figure}
\clearpage

\begin{figure}
\begin{center}
\includegraphics[height=8cm]{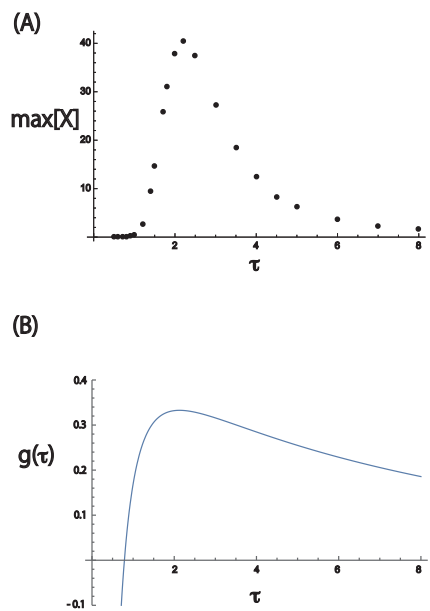}
\caption{(A) The resonant curve showing the maximum oscillatory amplitude, $max[X]$, with values of the delays $\tau$ as the tuning parameter. The parameters are set as the same as Fig.1; $a=0.01, b= - 2.0$ with the initial interval condition as $X(t) = 0.1 (-\tau \leq t \leq 0)$. (B) The plot of the function $g(\tau)$ given in (\ref{resoW}) with $a=0.01, b= - 2.0$}.
\label{resonant point}
\end{center}
\end{figure}

\begin{figure}
\begin{center}
\includegraphics[height=10cm]{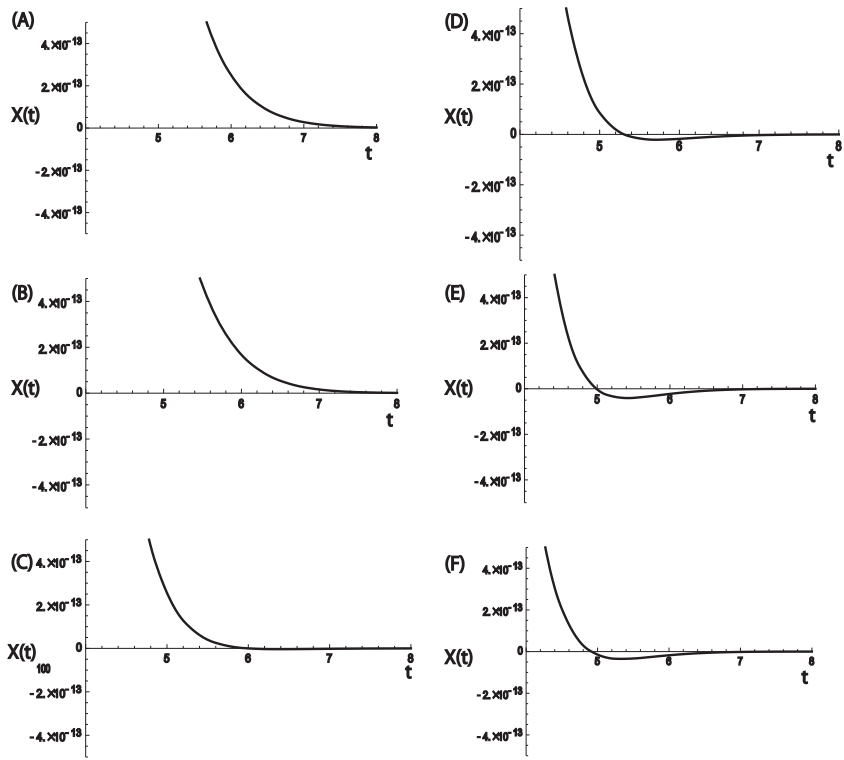}
\caption{Representative examples of change from monotonic to non-monotonic approach to the stable origin. The delay for this onset of the oscillation is approximated at $\tau_o \approx  0.368$ with $a = 0.01, b = - 1.0$.
The values of the delays $\tau$ are (A)$0.34$, (B)$0.35$, (C)$0.36$, (D)$0.37$, (E)$0.38$, (F)$0.39$.}
\label{critical}
\end{center}
\end{figure}
\clearpage

\begin{figure}[h]
\begin{center}
\includegraphics[height=6cm]{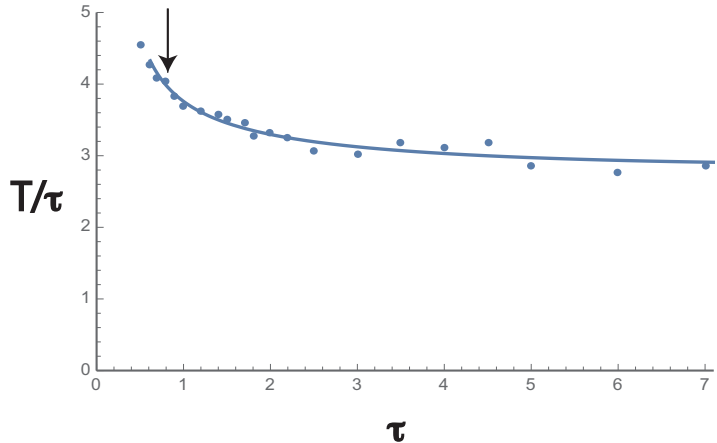}
\caption{The ratio of the period $T$ of the dominant oscillation mode to the delay $\tau$. The parameters are set as the same as Fig.1; $a=0.01, b= -2.0$ with the initial interval condition as
$X(t) = 0.1 (-\tau \leq t \leq 0)$. The estimated values of the critical delay $\tau_c  = 0.788$ is indicated by the 
arrow at which $T_c  \approx 4.0 \tau_c$. The solid line is the approximation given by (\ref{Wperiod}).}
\label{period}
\end{center}
\end{figure}

\section{Discussion}

In this paper, we have investigated the delay differential equation (\ref{dr2}) both analytically and numerically. Its solution is semi-analytically tractable in the sense it can be expressed in the form of equation  (\ref{solwf3}) formally and approximated in the form of equation (\ref{solwf4}) using the Lambert $W$ function. The $W$ function has been used for the stability analysis\cite{shinozaki} and for a dynamical approximation for a fixed delay\cite{pusenjak2017}. Our analysis here, however, extends its usage and captures dynamics with varying delays reasonably well, which are normally considered as difficult with investigations of general delayed dynamical systems.
We also confirmed numerically that the asymptotic stability of the origin is kept even with a larger delay due to the quadratic exponential factor. 

At the same time, transient oscillations are observed. Notably, the amplitude changes with changing delay showing a resonant behavior with respect to the delay. Also, as mentioned, the estimation of the value of the delay at the resonant point is achieved by the use of the $W$ function. The resonant behaviors have been observed with the simpler equation (\ref{dr}) without an exponential factor. There, however, it is the peak of the power spectrum that went through the maximum, showing ``regular'' oscillations. In this sense, even though equations (\ref{dr}) and (\ref{dr2}) both show transient oscillation behavior, the resonant phenomena are different in nature. It is also different from delay-induced transient oscillation (DITO)\cite{milton_mmnp,pakdamanetal1998a}. The DITO is a solution of a coupled set of delay differential equations. With DITO, the transient oscillations do not get suppressed as in our model, but oscillatory behaviors keep a prolonged duration of oscillation with increasing delay.

Similar transient oscillations are observed with some other equations 
in the form of equation (1). Our preliminary numerical investigation using a monotone decreasing (``negative feedback'') and Mackey--Glass (``mixed feedback'') functions\cite{mackeyglass1977,glass1988,glassmackey1988} show transient oscillations. The detailed nature of these equations and associated behaviors are left for future studies.
\vspace{2em}

\noindent
{\bf Acknowledgments}


The authors would like to thank the useful discussions with Prof. Hideki Ohira and the members of his research group at Nagoya University. This work was supported by the ``Yocho-gaku" Project sponsored by Toyota Motor Corporation, JSPS Topic-Setting Program to Advance Cutting-Edge Humanities and Social Sciences Research Grant Number JPJS00122674991, JSPS KAKENHI Grant Number 19H01201, and the Research Institute for Mathematical Sciences,
an International Joint Usage/Research Center located in Kyoto University.

\section*{Appendix}

In this appendix, we tabulate the values of the $W$ function that are used in and related to the main text of this paper (Table 1 and 2). The values 
are from Mathematica(R), ver. 13. We also plot them in Figs. 6 and 7. 
We note from these that, in the range of parameters we discuss our model, we have the case that
the following holds.
\vspace{1em}

$Max[Re[W_k(x)]] = Re[W_0(x)] = Re[W_{-1}(x)]$,

and

$Min[|Im[W_k(x)]|] = Im[W_0(x)] = -Im[W_{-1}(x)]$.

\begin{table}[htbp]
\begin{center}
\begin{tabular}{|c|c|c|c|c|c|c|c|c|c|c|c|c|c|c|c|c|c|c|c|c|c|} \hline
$k$ & 0.2 & 0.5 & 0.8 & 1.0 & 1.5 & 2.0 & 5.0  & 10.0 \\  \hline
-10 & -4.98119 & -4.06499 & -3.59655 & -3.37504 & -2.97557 & -2.69648 & -1.88488 & -1.56667\\ 
-9 & -4.86717 & -3.95077 & -3.48225 & -3.26071 & -2.86119 & -2.58208 & -1.77043 & -1.45221\\
-8 & -4.73849 & -3.82181 & -3.35317 & -3.13159 & -2.7320 & -2.45285 & -1.64115 & -1.32293\\
-7 & -4.59080 & -3.67370 & -3.20491& -2.98326 & -2.58359 & -2.30439 & -1.49263 & -1.17441\\
-6 & -4.41749 & -3.49977 & -3.03074 & -2.8090 & -2.40920 & -2.12994 & -1.31811 & -0.999913\\
-5 & -4.20779 & -3.28902 & -2.81962 & -2.59774 & -2.19774 & -1.91839 & -1.10652 & -0.788393\\
-4 & -3.94219 & -3.02150 & -2.55143 & -2.32931 & -1.92901 & -1.64954 & -0.837777 & -0.519876\\
-3 & -3.57946 & -2.65445 & -2.18303 & -1.9604 & -1.55973 & -1.2802 & -0.46942 & -0.15237\\
-2 & -3.00248 & -2.06355 & -1.58892 & -1.36578 & -0.965451 & -0.687324 & 0.114606 & 0.426424\\
-1 & -0.94422 & -0.319003 & 0.010835 & 0.16922 & 0.458766 & 0.664004 & 1.27425 & 1.51876\\
0 & -0.94422 & -0.319003 & 0.010835 & 0.16922 & 0.458766 & 0.664004 & 1.27425 & 1.51876\\
1 & -3.00248 & -2.06355 & -1.58892 & -1.36578 & -0.965451 & -0.687324 & 0.114606 & 0.426424\\
2 & -3.57946 & -2.65445 & -2.18303 & -1.9604 & -1.55973 & -1.2802 & -0.46942 & -0.15237\\
3 & -3.94219 & -3.02150 & -2.55143 & -2.32931 & -1.92901 & -1.64954 & -0.837777 & -0.519876\\
4 & -4.20779 & -3.28902 & -2.81962 & -2.59774 & -2.19774 & -1.91839 &-1.10652 & -0.788393\\
5 & -4.41749 & -3.49977 & -3.03074 & -2.8090 & -2.40920 & -2.12994 & -1.31811 & -0.999913\\
6 & -4.59080 & -3.67370 & -3.20491 & -2.98326 & -2.58359 & -2.30439 & -1.49263 & -1.17441\\
7 & -4.73849 & -3.82181 & -3.35317 & -3.13159 & -2.7320 & -2.45285 & -1.64115 & -1.32293\\
8 & -4.86717 & -3.94952 & -3.48225 & -3.26071 & -2.86119 & -2.58208 & -1.77043 & -1.45221\\ 
9 & -4.98119 & -3.95077 & -3.59655 & -3.37504 & -2.97557 & -2.69648 & -1.88488 & -1.56667\\
10 & -5.08353 & -4.16749 &-3.69911 & -3.47763 & -3.07819 & -2.79913 & -1.98757 & -1.66936\\
\hline
\end{tabular}
\end{center}
\caption{Real part of the $W_k (\hat{b}\tau)$ for $-10 \leq k \leq 10$ and $\tau =\{0,2, 0.5, 0.8, 1.0,1.5,2.0,5.0,10\}$ with $a=0.01, b = -2.0$. }
\label{wreal}
\end{table}

\begin{figure}[h]
\begin{center}
\includegraphics[height=16.4cm]{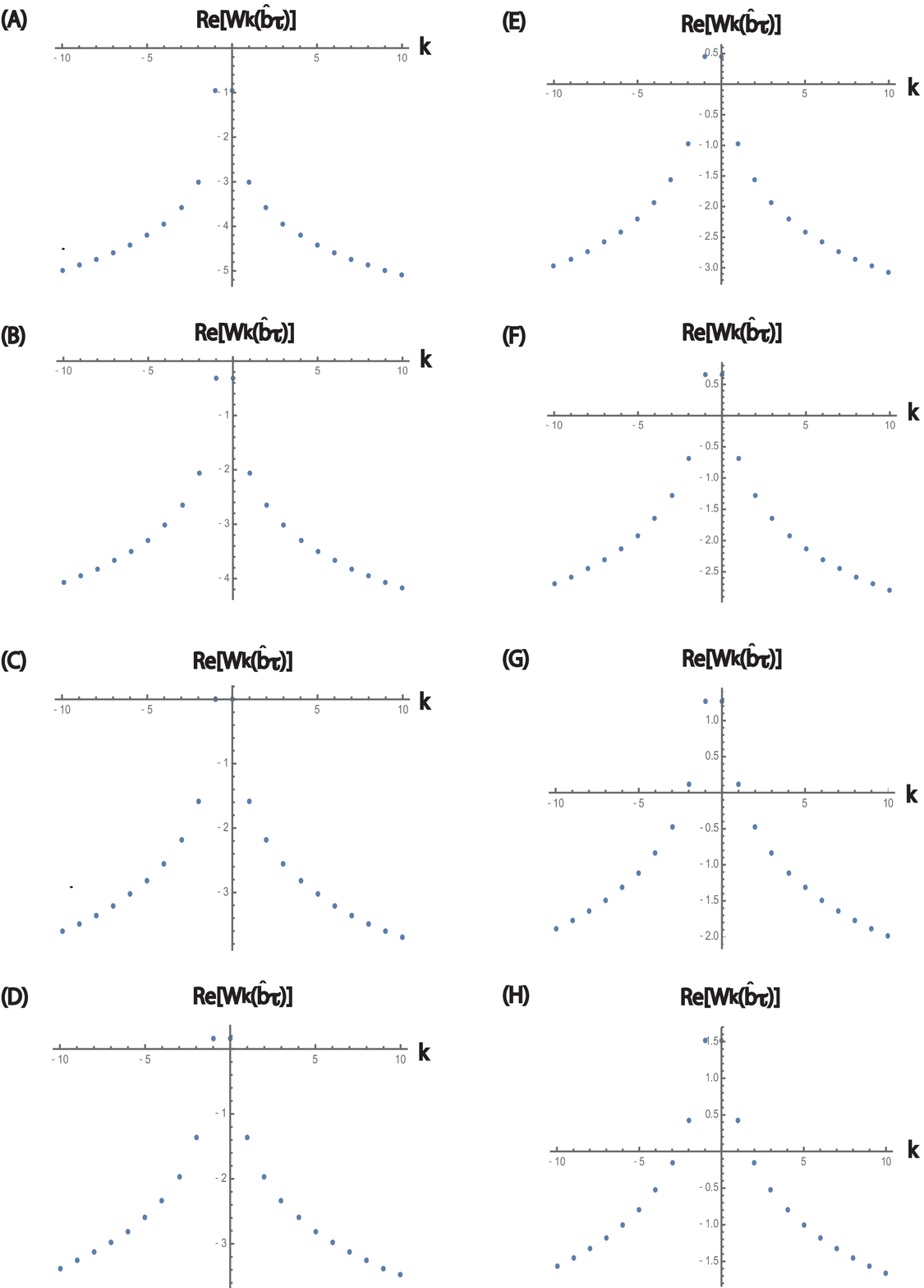}
\caption{Plots of Table 1. The values of the delays $\tau$ are (A)$0.2$, (B)$0.5$, (C)$0.8$, (D)$1.0$, (E)$1.5$, (F)$2.0$, (G)$5$, (H)$10$.}
\label{gwreal}
\end{center}
\end{figure}

\begin{table}[htbp]
\begin{center}
\begin{tabular}{|c|c|c|c|c|c|c|c|c|c|c|c|c|c|c|c|c|c|c|c|c|c|} \hline
$k$ & 0.2 & 0.5 & 0.8 & 1.0 & 1.5 & 2.0 & 5.0  & 10.0 \\  \hline
-10 & -58.0338 & -58.0496 & -58.0576 & -58.0614 & -58.0683 & -58.0731 & -58.0870 & -58.0925\\
-9 & -51.7425 & -51.7601 & -51.7691 & -51.7734 & -51.7811 & -51.7865 & -51.8021 & -51.8083\\
-8 &-45.4492 & -45.4692 & -45.4795 & -45.4844 & -45.4931 & -45.4992 & -45.5171 & -45.5240\\
-7 & -39.1532 & -39.1764 & -39.1883 & -39.1939 & -39.2041 & -39.2112 & -39.2319 & -39.2400\\
-6 & -32.8531 & -32.8807 & -32.8948 & -32.9016 & -32.9137 & -32.9221 & -32.9467 & -32.9564\\
-5 & -26.5463 & -26.5804 & -26.5979 & -26.6062 & -26.6212 & -26.6316 & -26.6621 & -26.6740\\
-4 & -20.2279 & -20.2724 & -20.2953 & -20.3061 & -20.3257 & -20.3394 & -20.3793 & -20.3949\\
-3 & -13.8849 & -13.9491 & -13.9823 & -13.9980 & -14.0264 & -14.0463 & -14.1039 & -14.1264\\
-2 &-7.47189 & -7.58847 & -7.64917 & -7.67794 & -7.72972 & -7.76570 & -7.86855 & -7.9079\\
-1 &-0.406786 & -1.33649 & -1.57766 & -1.67168 & -1.81798 & -1.90602 & -2.11338 & -2.17949\\
0  & 0.406786 & 1.33649 & 1.57766 & 1.67168 & 1.81798 & 1.90602 & 2.11338 & 2.17949\\
1 & 7.47189 & 7.58847 & 7.64917 & 7.67794 & 7.72972 & 7.76570 & 7.86855 & 7.9079\\
2 & 13.8849 & 13.9491 & 13.9823 & 13.9980 & 14.0264 & 14.0463 & 14.1039 & 14.1264\\
3 & 20.2279 & 20.2724 & 20.2953 & 20.3061 & 20.3257 & 20.3394 & 20.3793 & 20.3949\\\
4 & 26.5463 & 26.5804 & 26.5979 & 26.6062 & 26.6212 & 26.6316 & 26.6621 & 26.6740\\
5 & 32.8531 & 32.8807 & 32.8948 & 32.9016 & 32.9137 & 32.9221 & 32.9467 & 32.9564\\
6 & 39.1532 & 39.1764 & 39.1883 & 39.1939 & 39.2041 & 39.2112 & 39.2319 & 39.2400\\
7 & 45.4492 & 45.4692 & 45.4795 & 45.4844 & 45.4931 & 45.4992 & 45.5171 & 45.5240\\
8 & 51.7425 & 51.7601 & 51.7691 & 51.7734 & 51.7811 & 51.7865 & 51.8021 & 51.8083\\
9 & 58.0338 & 58.0496 & 58.0576 & 58.0614 & 58.0683 & 58.0731 & 58.0870 & 58.0925\\
10 & 64.3238 & 64.338 & 64.3452 & 64.3487 & 64.3549 & 64.3592 & 64.3718 & 64.3767\\
\hline
\end{tabular}
\end{center}
\caption{Imaginary part of the $W_k (b\tau)$ for $-10 \leq k \leq10$ and $\tau =\{0,2, 0.5, 0.8, 1.0,1.5,2.0,5.0,10\}$ with $a=0.01,b = -2.0$. }
\label{wimag}
\end{table}

\begin{figure}[h]
\begin{center}
\includegraphics[height=16.4cm]{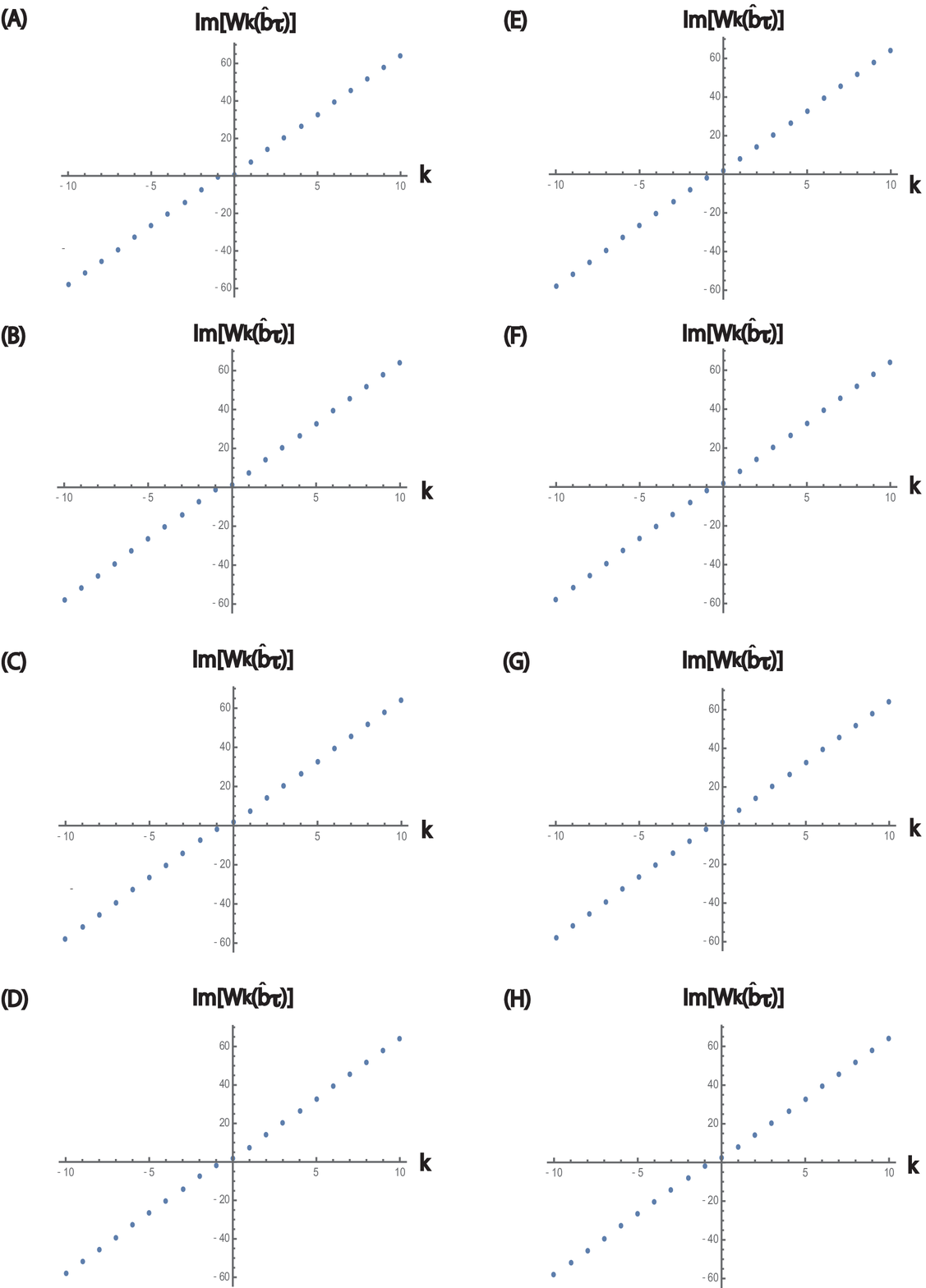}
\caption{Plots of Table 2. The values of the delays $\tau$ are (A)$0.2$, (B)$0.5$, (C)$0.8$, (D)$1.0$, (E)$1.5$, (F)$2.0$, (G)$5$, (H)$10$. }
\label{gwimag}
\end{center}
\end{figure}

\end{document}